\documentclass[runningheads]{llncs}
\usepackage{graphicx}
\usepackage{draftwatermark}
\SetWatermarkText{PREPRINT}
\SetWatermarkScale{5}
\begin{document}
\title{RoboCup Junior in the Hunter Region: \\Driving the Future of Robotic STEM Education}
%
\titlerunning{RoboCup Junior, Hunter Region, Australia}
%
\author{Aaron S.W. Wong\inst{1,2,3} \and
Ryan Jeffery\inst{2} \and
Peter Turner\inst{1,2,4} \and 
Scott Sleap\inst{5} \and
Stephan K. Chalup\inst{1,2}}
\authorrunning{A.S.W. Wong et al.}
%
\institute{Newcastle Robotics Laboratory, The University of Newcastle \and Faculty of Engineering and Built Environment, The University of Newcastle \and Faculty of Science, The University of Newcastle \and Tribotix Pty. Ltd. \and Regional Development Australia -  Hunter \\
\email{aaron.wong@newcastle.edu.au}}
\maketitle              
\begin{abstract}
RoboCup Junior is a project-oriented educational initiative that sponsors regional, national and international robotic events for young students in primary and secondary school. It leads children to the fundamentals of teamwork and complex problem solving through step-by-step logical thinking using computers and robots. The Faculty of Engineering and Built Environment at the University of Newcastle in Australia has hosted and organized the Hunter regional tournament since 2012. 
This paper presents an analysis of data collected from RoboCup Junior in the Hunter Region, New South Wales, Australia, for a period of six years 2012-2017 inclusive. Our study evaluates the effectiveness of the competition in terms of geographical spread, participation numbers, and gender balance. We also present a case study about current university students who have previously participated in RoboCup Junior.

\end{abstract}
\section{INTRODUCTION}

The National Innovation and Science Agenda \cite{Commonwealth2016} is the Australian government's initiative to improve and promote technologically related fields of commercial, industrial, and technical skill development for Australian citizens with a component focusing on the preparation of young school-aged students for studies related to the fields of Science, Technology, Engineering, and Mathematics (STEM). This component has been recognised as the next step in the evolution of the Australian education system. It is understood that the next generation of graduating students will have to be ``STEM-ready'' to cope with the challenges of a future technologically advanced and internationally competitive workforce. 

Educating a STEM-ready workforce is not a trivial task as there are several behavioural factors inherent to modern Australian culture which impede this goal. These include fear of failure \cite{Michou2014} (of both students and their teachers), and mathematics anxiety \cite{wigfield1988}. Students in general should be encouraged at an early stage to take intellectual risk in order to gain life-skills that could be of value for a STEM-related career path \cite{AIG2015}. 
For students, pathways into a STEM related field can be increased by diversifying opportunities and options, so that there is a higher probability of attracting their attention.
They could start, for example, with programming websites or building electronic components. In order to help students to access STEM, it is important to explain practical applications and give STEM a purpose. It also should be made clear that mathematics consists of many different disciplines and that they may require very different ways of thinking. There are different options within STEM and similarly within the general area of mathematics, e.g., not everyone who is talented in geometry and visualisations may also be good at memory and number tasks. With these critical thoughts in mind, a STEM-ready workforce has the potential to advance many different aspects of technology. 

One aspect of an advanced technology can be found in the field of robotics. Many STEM-related fields and skills are required to develop an autonomous robot. These include sound knowledge of the fundamental concepts of science and mathematics as taught at school. These are also the basis to a successful development of the practical skills required by professional engineers. 

Similarly as the Personal Computer (PC) and the Internet had revolutionary impact on our culture, now mobile devices and robots are predicted to be basis of the next technological explosion. Hence, it is important to encourage young students to consider gaining skills in professions related to robotics or to pursue a other career paths that also can lead to a technologically skilled future, i.e., a ``STEM-ready'' future. For the Hunter region in New South Wales (NSW), Australia, RoboCup Junior \cite{RobocupJuniorWeb} is the only annual robotics event that targets young students to actively compete and perform as a team. RoboCup Junior is a project-oriented educational initiative that sponsors regional, state, national and international robotic events for young students, with the goal to encourage the next generation to pursue and take an interest in scientific and technological fields. 

In this paper we will investigate the following question: How can the current society be prepared for a sustainable STEM-minded future?  How can a community engagement project such as RoboCup Junior Hunter contribute and how can its success be measured?

The subsequent sections show how RoboCup Junior served as a platform for community engagement and for promoting STEM in the Hunter region. By detailing demographic information about the Hunter region it highlights the importance and impact of RoboCup Junior. The results section provides quantitative measures derived from data collected from RoboCup Junior events over the past 6 years. Then a case study of current University of Newcastle students who have previously participated in the competition is presented. The penultimate section discusses the importance and roles of several stakeholders that collaboratively supported RoboCup Junior in the Hunter region and how this led to one of the most successful regional initiatives of its kind in Australia.

\section{ROBOCUP JUNIOR IN THE HUNTER REGION}

\begin{figure}[h]
\centering
\includegraphics[width=\textwidth]{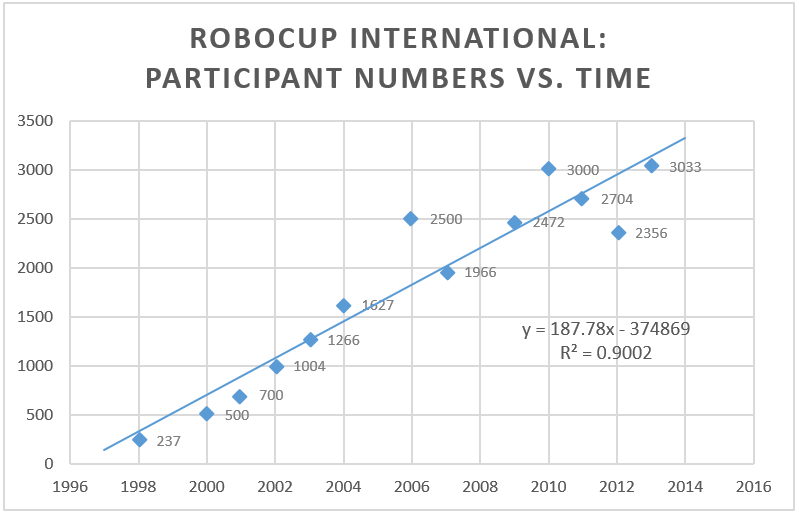}
\caption{The number of participants who have attended the RoboCup International tournament since its inception \cite{Veloso2012}: There is a strong linearly increasing trend ($R^2 > 0.90$) over time with the number of people globally participating in this tournament.}
\label{fig:1}
\end{figure}

RoboCup (est. 1997) is an international competition that fosters research in robotics \cite{Kitano1997}, and the advancement of artificial intelligence within a competitive environment. RoboCup has seen a globally increasing trend in the past decade, see Fig. \ref{fig:1}. The goal of RoboCup, in the near future, is to have designed and programmed a team of bipedal humanoid autonomous soccer playing robots, to win against the human world champion soccer team \cite{Veloso2012}. This goal is yet to mature, and may require some generations of research to achieve, and hence we have RoboCup Junior; the establishment of the next generation of ``technologists'', with a focus on robotics. 

RoboCup Junior is designed to introduce primary and secondary school-aged children to the fundamentals of teamwork and complex problem solving by employing step-by-step logical, rational processes using computers (robots) as a tool to complete a set task. The main objective of RoboCup Junior is to encourage the next generation to pursue and take an interest in scientific and technological fields; to cultivate their interests through a hands-on approach in robot design and creation using platforms including, but not limited to, Lego Mindstorm educational kits. Students are invited to compete in three distinct disciplines; soccer, dance, and rescue.  

With the RoboCup Junior initiative, it is possible to create an environment of light-mindedness, experimentation, fun and teamwork that inspires and educates students to expand their horizon through STEM experiences. In this context, there are countless opportunities to establish links to other associated STEM disciplines. It is important that students feel respected as individuals and that they have, at an early stage, access to demonstrations and practical hands-on aspects of STEM careers in a broad manner where they can explore their own career goals. This is one of the key and defining ideals of RoboCup Junior in the Hunter region, NSW.

The Hunter region, NSW, resides approximately two hours north of Sydney and has a substantial rural demography as well as large urban population centres in Newcastle and Lake Macquarie. Although, the Hunter region has been noted to be an innovation hub, e.g. Newcastle as ``Smart City'' \cite{hunterbusinessreview_2016}, where new smart technologies are being developed as applied solutions to the problems of the world today, the general population includes negatively skewed low social economic status (SES) indicators, when compared to the state's capital, Sydney. For example, Higher School Certificate (HSC) completion rates in the Cessnock Area, within the Hunter region, are with 44\% substantially lower than the NSW average of 75\% \cite{CesnockCC2015}. These low-SES indicators have led the organisers of RoboCup Junior in the Hunter to follow an approach that maintains the core rules of the competition while incorporating additional coaching to promote participation. This allows children to have an attitude of ``having-a-go'', to have fun, and to enjoy themselves while avoiding anxieties associated with STEM subjects and while subconsciously having a positive experience with STEM and gaining important skills required for a potential career path in STEM.

A career path in STEM does not require direct entry into a university degree, as there are many different pathways into a STEM-related career. However, traditional entry into a university STEM-related degree has generally been perceived to be the fastest arrangement to refine skills, achieve, advance and progress in a competitive STEM workforce. For this reason, the following results section presents information obtained in a case study of currently enrolled students (with their permission) who were past participants of RoboCup Junior, Hunter. This case study together with quantitative results recorded over the past six years corroborates the view that the Hunter RoboCup Junior initiative had substantial positive impact on driving STEM education in the Hunter Region.

\section{RESULTS}

\subsection{QUANTITATIVE ANALYSIS}

\begin{figure*}[htb!]
\centering
\includegraphics[width=\textwidth]{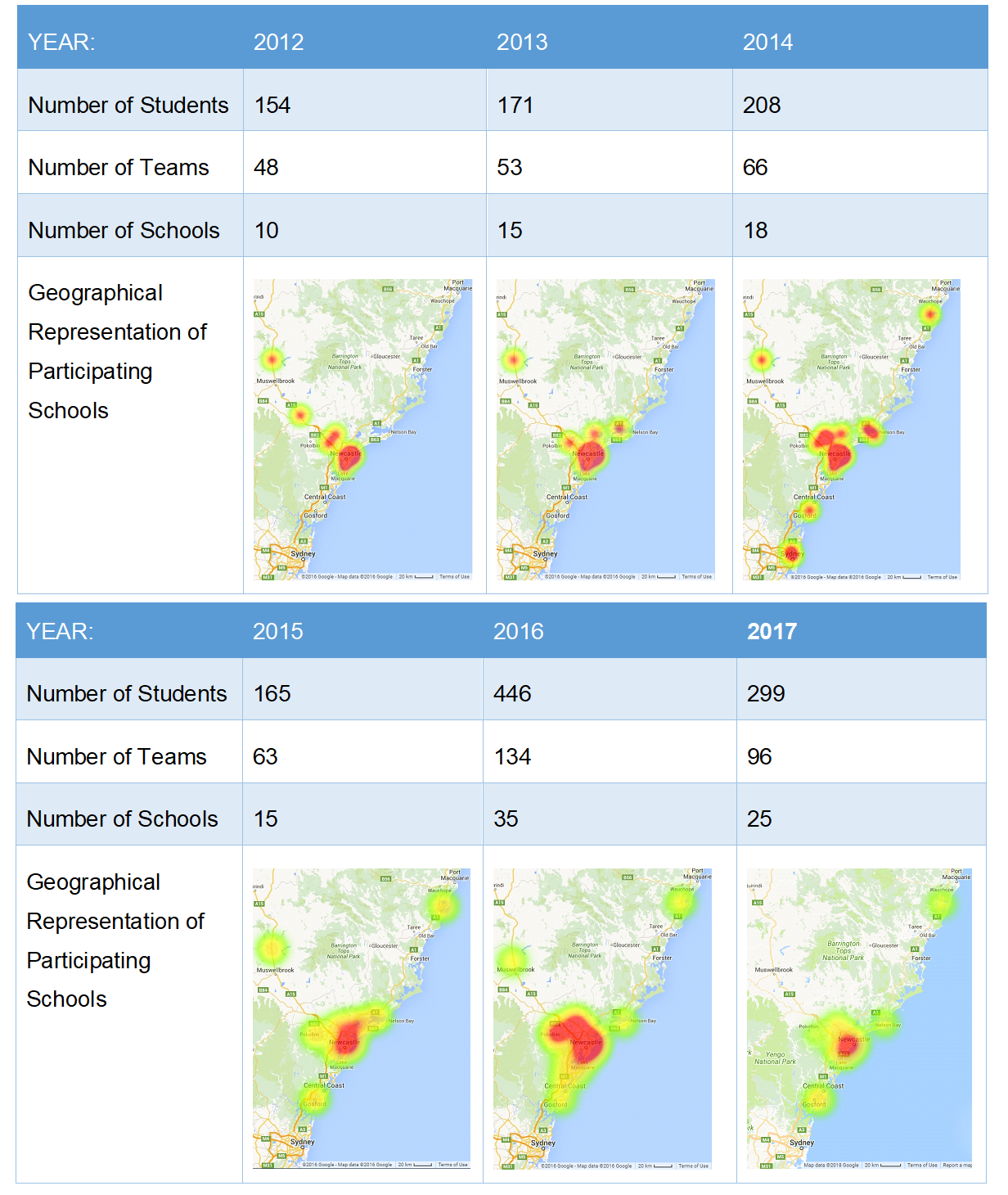}
\caption{Development of RoboCup Junior in the Hunter Region over the past six years (2012-2017 inclusive). There was a general increase in the number of students, teams, and schools. The spatial coverage of the participating schools entering the competition also increased.}
\label{fig:2}
\end{figure*}

Over the past six years, 2012 to 2017 (see Fig.~\ref{fig:2}) there has been a general growth in the number of students, teams, as well as schools, with a total of 1443 student participants in the Hunter Region RoboCup Junior competition. In 2015 the ME program (Section~\ref{sec:ME}) was not able to support the competition as usual. While in 2017, the date of the competition was significantly earlier then previous years. These factors caused a temporary decline in participation. 

Students were as young as 8 years of age in their 3rd school year, and the oldest participating students were up to 18 years of age. The majority of the participating students was aged 14 and 15 (in school years 8 and 9). About 22\% of the 1443 participating students were female (2012=43, 2013=18, 2014=46, 2015=32, 2016=106 and 2017=71). For an extra-curricular school activity this fraction can be considered to be relatively high, given that only 13\% of all engineers are female~\cite{engineersaustralia}. As the geographical representation of participating schools in Fig. \ref{fig:2} shows the geographical distribution of schools participating in the competition increased as the initiative matured over time.  Participating schools were from the Central Coast in the South up to Camden Haven in the North. In addition  some schools travelled up from the Sydney Region to attend the Hunter tournament in 2014.

\begin{figure}[h]
\includegraphics[width=\textwidth]{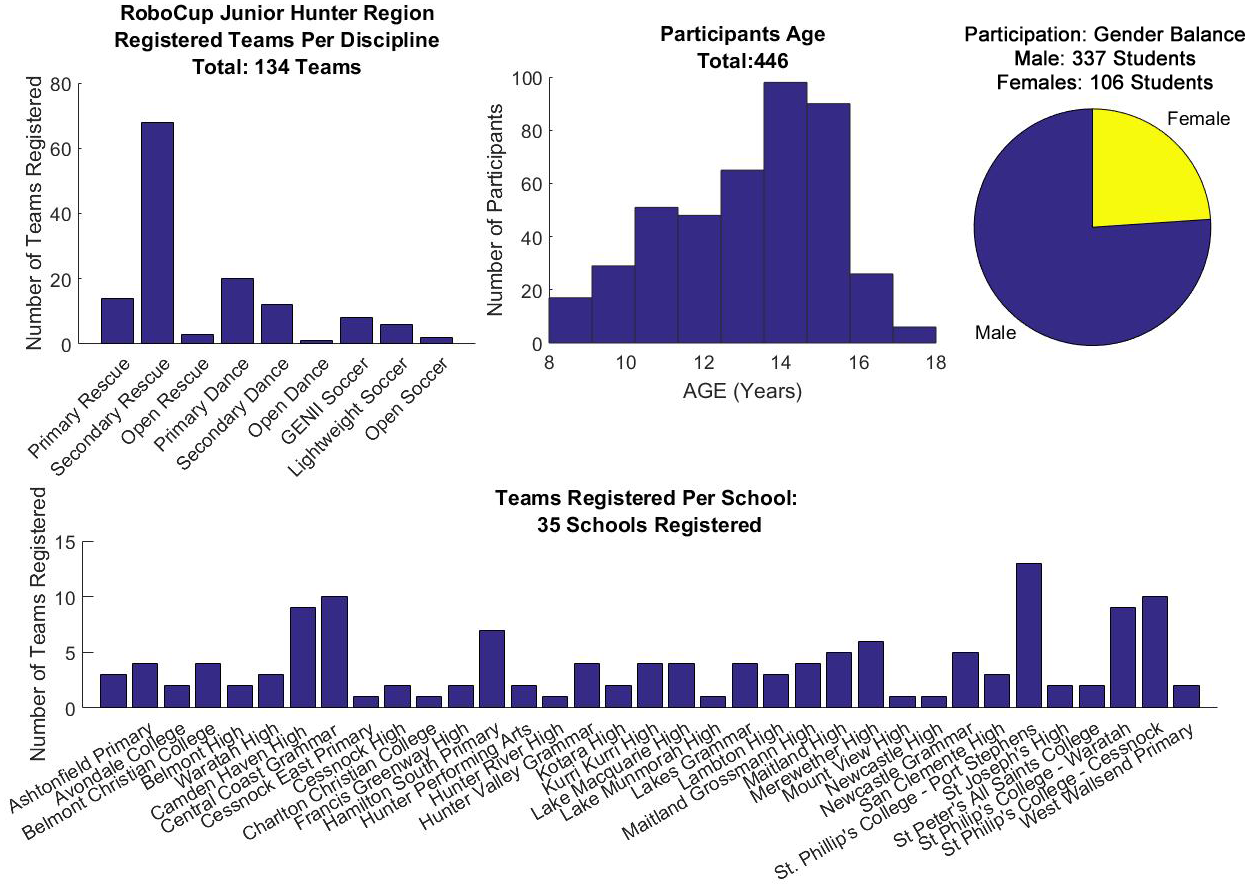}
\caption{Distributions of registered participants for 2016, RoboCup Junior, Hunter Region: Top Left, a histogram representing the number of teams registered per discipline. Top Middle, a histogram representing the number of participants by age. Top Right, gender distribution for the competition. Bottom, a histogram representing teams registered by school.}
\label{fig:3}
\end{figure}

With respect to the disciplines offered at RoboCup Junior, the local Hunter region competition comprises all available disciplines that are currently accessible at both state and national tournaments. As a result, the discipline participation distribution follows the identical ranking in proportion with the difficulty of discipline. The ranking from easiest to hardest is as follows; rescue, dance, and soccer. The data from our largest local competition, in 2016, is shown in Fig.~\ref{fig:3}.

RoboCup Junior rescue consists of three sub-disciplines; primary, secondary, and open. Simplest sub-category is primary rescue, predominantly for students who are currently in primary school, while secondary rescue is for participants who are attending high school, or are between 13 and 18 years old. Advanced participants who have participated in at least two tournaments of rescue in previous years compete in open rescue. This hierarchy of sub-categories allows not only for an increase in difficulty, but for different challenges for the different categories of ages. The rescue discipline is very structured, and the task is set by the rules of the competition. This structure was the motivating factor that drove its popularity, with 80 (14 primary, 68 secondary) teams registered to participate in this discipline in 2016.

Like rescue, RoboCup Junior dance also has three sub-categories; primary, secondary and open, with similar age restrictions. This category leaves many open-ended opportunities and flexibility that allow students to explore and experiment with different implementations for their performance. This openness makes it somewhat further challenging when compared to the rescue discipline. Hence, the number of participants in RoboCup Junior dance is smaller than in the rescue category. The number of teams participating in each sub-category is fairly distributed, with approximately 20 teams for primary dance, and 14 teams for secondary dance, with one team for open dance.

Lastly, in RoboCup Junior soccer, the students are required to build and program a small robot team to autonomously play soccer. It is essential for the robots to autonomously adapt to the environment while playing soccer by the rules. This discipline is the utmost challenging of the disciplines offered at the local Hunter regional tournament. However, it is also the most rewarding of all disciplines, as it allows the students to program an autonomous agent, with a requirement for  multivariate control algorithm be used. The other disciplines can be, but are not necessarily, much simpler. Due to its perceived difficulty, participation in the soccer discipline has dwindled throughout the history of our local tournament. RoboCup Junior soccer, like RoboCup Junior rescue, consists of three sub-disciplines; GENII, lightweight, and open. The sub-category of GENII only allows Lego Mindstorm NXT or EV3 Robots to compete, whereas lightweight allows modified self-built robots to be used. This includes Arduino based hardware under a certain weight and size limit. The open sub-discipline allows any hardware to be used in the build of robots, which could be of any weight, but within a size limitation. For 2016, we saw the largest cohort, with 18 teams in soccer discipline total (9 GENII, 7 Lightweight, and 2 Open).



\subsection{A CASE STUDY}
A report released in 2015 by the Australian Industry Group, explains a set of key recommendations that can be implemented in order to further encourage school students to study and explore future careers in STEM \cite{AIG2015}. A particular key recommendation highlights the need for teachers and schools to be further supported in harnessing students' interest, which is the primary aim of RoboCup Junior and the number of community partnerships involved. Burgher et al. \cite{burgher2015implementation} suggest that a hands-on and practical approach to education, results in students becoming more aware of the conceptual theory rather than students being traditionally educated in the format of lectures and traditional classroom exercises. Exploring this suggestion this study has taken the shape of a questionnaire to current university students (n=3), who have competed in the tournament in the past. The purpose of investigating this qualitatively was to understand what key factors associated with their participation in RoboCup Junior led the students to develop a deeper appreciation of STEM and finally resulted in their decision to pursue a career in STEM. To begin, all participants had indicated in the questionnaire that they had been encouraged to participate in RoboCup Junior by their schools and teachers. This exemplifies how critical and significant partnerships formed between the organising committee and schools are. Three prevalent themes in the responses consisted of the following areas; Problem solving, Conceptual thinking and Rewarding.
\paragraph{Problem Solving:}
Responses indicated that students had found the nature of the problems presented in RoboCup Junior to be of a ``broad nature'', which provided further incentive to aid the design of a solution they were presented with during the competition. A critical skill that has been cited is that students had to assess the abstract problem independently which leads into the second theme.
\paragraph{Conceptual thinking:}
In alignment with Burgher et al. \cite{burgher2015implementation} RoboCup Junior being a practical activity, allowed students to gain an additional direct approach to conceptual learning. Results from the questionnaire indicated that the critical skills gained were of a separate nature to a school curriculum. Students also indicated that open-ended problems allowed them to focus on concepts rather than on textbook knowledge. Students also suggested that as a result of it being separate from a school curriculum it allowed focus to bridge gaps between abstract and practical nature.
\paragraph{Rewards:}
Within RoboCup Junior, students are encouraged to use technology to solve a given problem. The nature of responses to the questionnaires indicates that participants find the solutions to be the most rewarding and therefore one of the most encouraging aspects. As STEM is centrally focused around problem solving, students who experienced this during the competition found it further encouraging to seek employment within a career that offers that same sense of reward for solving a broad problem.

\section{DISCUSSION: PARTNERSHIPS}

The future of STEM in the Hunter region is deemed important by many STEM-related stakeholders of the local region. Several factors that have contributed to the success of the tournament are associated with the partnerships that have been developed with key stakeholders in the period 2012-2017. Some of these key relationships and their impact will be discussed in the following sub-sections. 

\subsection{THE UNIVERSITY OF NEWCASTLE, FACULTY OF ENGINEERING AND BUILT ENVIRONMENT}
The Faculty of Engineering and Built Environment of the University of Newcastle has continuously been the main stakeholder of RoboCup Junior in the Hunter Region since the project's re-inception in 2012. With the university acting as the host, the competition is held on campus at the university's gymnasium, The Forum. The university is also the key supplier of human resources to organise and manage the competition. The faculty has supported the competition with expertise in management, in the form of faculty administrative staff. The faculty manages aspects such as registrations, budget, and covered the majority of costs to run the competition.

Technical aspects of the tournament were administered by members affiliated with the NUbots, the Newcastle University RoboCup team. The NUbots are a senior RoboCup team and comprise several university students and academics~\cite{NUbots2018}. They competed in the Kidsize Humanoid League, and now in 2018, the TeenSize Humanoid League at RoboCup. They are part of the Newcastle Robotics Laboratory, situated in the Faculty of Engineering and Built Environment. The NUbots have participated in RoboCup since 2002. They became world champions in the Standard Platform League using the Aldebaran NAO Robots, in 2008, and were world champions in the 4-Four-Legged League in 2006 using the Sony AIBO robots. This internationally well-recognised team brings over a decade of robotics experience to the local RoboCup Junior Hunter region competition. NUbot members are members of the committee, deliver workshops, and play a crucial role on the Hunter Region competition day in roles such as technical refereeing and judging. Over the past decade, the faculty has had a strong interest in community engagement. With a particular interest in low social-economical-status areas, the faculty has deployed intensive training programs and funding for robotic kits at schools in areas such as, Raymond Terrace in Port Stephens,~NSW. 

\subsection{REGIONAL DEVELOPMENT AUSTRALIA (RDA) HUNTER – MANUFACTURING ENGINEERING (ME) PROGRAM}\label{sec:ME}

RDA – Hunter's ME Program has been a highly successful STEM outreach program that has delivered tangible outcomes in terms of student uptake of STEM-based subjects in upper secondary schools \cite{Sleap2014}.  The ME Program in the Hunter region has supported running of the RoboCup competition whenever possible during 2012- 2014 and 2016. 

In addition to supporting RoboCup Junior, the ME Program actively supports all aspects of STEM in the Hunter and has produced an innovative school curriculum which integrates the silos of STEM into a Year 9 and 10 elective subject (iSTEM), which was endorsed by BOSTES NSW in 2012. In 2017, there were over 100 schools across NSW teaching iSTEM, which includes robotics programs in a standard curriculum, which also includes RoboCup Junior preparation. As a result of the broader ME Program funding for local schools, it has delivered a substantial quantity of STEM equipment and training (e.g. professional learning for teachers and through the support of Robogals for schools). The hardware provided includes 3D printers, and of course, robotic kits that could be used as part of the RoboCup Junior competition. The 2016 ME Program has included a caveat for any school receiving Lego EV3 robots that they must compete in the RoboCup Junior, Hunter region competition. 
During 2016, there was a significant increase in the number of schools that received robotic kits as part of the ME program. During 2015-2016, over one hundred EV3 robots were provided to 22 local schools. As a consequence, there was a significant increase in registered participants for the local tournament in 2016. 

\subsection{ROBOGALS NEWCASTLE INITIATIVE}

Robogals is an international initiative aimed at promoting gender equality in the fields of STEM through the use of robots and robotic education. Volunteers of the initiative consistently visit different schools and perform their robots at local public events. In addition, they offer free short beginner classes in robotics in using the Lego NXT and EV3 Robots in many school classes. The local chapter of Robogals in the Hunter region is no exception. It consists of many enthusiastic individuals, who are always ready to assist and share when required. The local chapter of Robogals initiative has worked closely with the RoboCup Junior Hunter Region Competition, since the inception of the local chapter in 2013. Robogals have recently signed a Memorandum of Understanding with the ME Program and BAE Systems Williamtown and have been working with ME Program high school and their feeder primary schools. The ME Program also provided 10 EV3 Robots to complement their fleet of NXT units. Volunteers of Robogals have sat on the organising committee for RoboCup Junior, Hunter region, and have also assisted at the events with judging, and holding workshops on the competitions behalf while the competition was running. In addition, training material used to teach classes was shared between RoboCup Junior and Robogals, so that Robogals could concentrate on their goal to achieve gender equality in the fields of STEM. The Robogals initiative in the Hunter region is growing successfully. They have repeatedly reported that there are more schools on their waiting list then they can handle. The result of this partnership, and its growing success, can be seen in the increased female to male ratios (approximately 24\% in Figure \ref{fig:3}). It shows the number of females is relatively higher at RoboCup Junior when compared to the number of females enrolled in an engineering course at a later stage, e.g., at university level.

\subsection{TRIBOTIX}
Tribotix is a local robotics company in the Hunter region that sells and builds various robots, mostly for educational purposes. It has a strong interest in the success of RoboCup Junior. The director of Tribotix has personally been involved with the RoboCup Junior since its inception and has also mentored teams in local schools, using a different style of robots than the standard Lego platform. 
Tribotix also partners with the national RoboCup Junior Australia committee in the development of state-of-the-art robotic educational kits. This includes, e.g., the DARwIn-MINI, for future use in a possible new Rescue league and a small humanoid league for RoboCup Junior.
As more students start earlier with the competition, it would not be too long before these advancing students seek knowledge, information, and new hardware to fulfil their requirements. Tribotix has been a competent partner and helpful supplier throughout all years of the competition.

\subsection{COMMUNITY SPONSORSHIP AND MEMBERSHIP}
Community support was vital for running the event. Support was supplied in terms of funding obtained from community grants, such as Orica (2014), AGL (2015), Newcastle Coal Infrastructure Group (NCIG) (2015), Newcastle City Council (2016), and the Kirby Foundation (2017). Without this funding, the competition itself could not have happened, and therefore no success could have been achieved. Members of the general community represented by teachers and parents of the participants were involved in all aspects of organising the competition. The success of the students comes directly from interacting with their mentors, some of which advise and attend monthly organising meetings which allow us to hear feedback and to incorporate and implement suggestions to make the competition run smoothly. Members of the community are an important part of the Hunter Region RoboCup Junior organising committee and have steered the competition to its current successful state. We acknowledge Mr. Jason Flood, Chair of Local Committee (all years, excluding 2014), for his commitment  and extraordinary effort that added to the project's success.

\section{CONCLUSION}

With decreasing levels of participation in mathematics and science within Australian schools, winning students' interest in the fields of STEM has become an uphill battle. Nonetheless, for the local region of the Hunter, the RoboCup Junior competition gained outstanding success. 
The partnership of RoboCup Junior Hunter Region with the University of Newcastle, and other key stakeholders such as the RDA Hunter's ME program, stands as a project that will transform the landscape for STEM education in the future. Success of the project to this point is reflected by the increasing number of student participants, a growing geographical distribution, and an improvement of gender balance. In addition, qualitative evidence of the positive influence on students participating in RoboCup Junior explains what impact the competition can play on students' path to a STEM career.
 



%
%
%
\newpage
 \bibliographystyle{splncs04}
 \bibliography{RJCHR}

\end{document}